\documentclass[fleqn,10pt]{wlscirep}
\usepackage{graphicx}

\title{First order reversal curves and intrinsic parameter determination for magnetic materials; limitations of hysteron-based approaches in correlated systems}

\author[1,*]{Sergiu Ruta}
\author[2]{Ondrej Hovorka}
\author[3]{Pin-Wei Huang}
\author[3]{Kangkang Wang}
\author[3]{Ganping Ju}
\author[1,*]{Roy Chantrell}
\affil[1]{Department of Physics, The University of York, Heslington, York YO10 5DD, United Kingdom}
\affil[2]{Faculty of Engineering and the Environment, University of Southampton, Highfield, Southampton, United Kingdom}
\affil[2]{Seagate Technology,Fremont, CA}

\affil[*]{sir503@york.ac.uk}

\begin{abstract}
The generic problem of extracting information on intrinsic particle properties from the whole class of interacting magnetic fine particle systems is a long standing and difficult inverse problem.
As an example, the Switching Field Distribution (SFD) is an important quantity in the characterization of magnetic systems, and its determination in many technological applications, such as recording media, is especially challenging. Techniques such as the first order reversal curve (FORC) methods, were developed to extract the SFD from macroscopic measurements. However, all methods rely on separating the contributions to the measurements of the intrinsic SFD and the extrinsic effects of magnetostatic and exchange interactions.  We investigate  the underlying physics of the FORC method by applying it to the output predictions of a kinetic Monte-Carlo model with known input parameters. We show that the FORC method is valid only in cases of weak spatial correlation of the magnetization and suggest a more general approach.

\end{abstract}
\begin{document}

\flushbottom
\maketitle

\thispagestyle{empty}

\section*{Introduction}
Identification of the intrinsic properties of magnetic nanostructures is central to the development of applications in a wide range of topics: information storage, biomedicine, permanent magnet development, and many more. The parameter identification techniques are at the heart of large scale material characterisation to quantify the properties of nanoscopic constituents of materials. For example, the optimisation of magnetic granular materials for the current and future hard disk drive technologies, such as  heat assisted magnetic recording (HAMR), or the synthesis of magnetic nanoparticles for molecular sensing and detection, imaging, and cancer therapy in biomedicine relies on the possibility of efficient and accurate identification of the physical properties of billions of magnetic nanoparticles, which requires analysis in a high dimensional parameter space and the employment of a statistical approach. 

 In such cases, direct measurements targeting individual particles become inefficient and infeasible. Instead, an indirect approach based on relating theoretical models to macroscopic experimental data and identifying the model parameters from the optimal fit becomes the most viable approach. This inverse problem solving methodology relies on the availability of a realistic model capable of i) reliably representing the physics of elementary constituents of a physical system, ii) accurately reproducing the macroscopic measurement data (forward problem) and iii) understanding the uniqueness properties of inverse solutions of the model, i. e. whether the identified parameter set is the only set allowing the model to accurately reproduce the measurement data \cite{Tarantola2006, bertero1998introduction, sarvas1987basic, tarantola2005inverse}. Unfortunately, inverse problems are often ill-posed and entire manifolds of parameters allow the models to reproduce the measurement data, which effectively translates to a significant error of the parameter identification. Such errors can only be reduced by providing new information from independent measurements. As a result, the development of identification techniques for materials characterisation remains a challenging problem.

In this work we consider the exemplar problem of identification of the switching field distribution (SFD) in magnetic particulate and granular systems. The SFD carries information about the intrinsic conditions for magnetisation reversal of individual magnetic particles and, for example, is a crucial characteristic determining the quality of high density magnetic media for current and future hard disk technologies, such as based on the bit patterned media or granular materials considered for HAMR \cite{Yang2015,Weller2014a}. This problem is especially challenging due to the strength and complexity of the interparticle interactions.
%
%
The recently developed identification schemes to extract the SFD based on the inverse problem solving approach include applications to assemblies of magnetic nanoparticles\cite{Rut}, and $\Delta H (M,\Delta M)$ method for granular materials relevant in magnetic recording \cite{Liu2008a,Hovorka2010a,Hovorka2012,Pisana2013} \cite{Hauet2014,Tabasum2013,Wang2008a}, which was motivated by the earlier $\Delta H_c$ methodology  \cite{Tagawa1991}. Simpler approaches to identify the SFD with variable degree of consistency were based on the differentiation of  hysteresis loop `de-sheared' to remove the contribution from magneto-static interactions \cite{Pike1999}, methods based on Preisach models \cite{Mayergoyz1991}, and methods based on analysing the transformed first order reversal curves (FORC), i.e. magnetisation curves generated by reversing the external magnetic field starting from a point on a major hysteresis loop branch (Fig. \ref{fig1}). 
 The FORC methods are equivalent to the classical Preisach modelling if the measurement data display microscopic memory of states of magnetic particles after the external field excursion (wiping-out property) and a minor hysteresis loop congruency \cite{Mayergoyz1985, Stancu2003}.

The FORC methods are used broadly as a tool for qualitative, and in some cases quantitative, description of general magnetic characteristics of magnetic systems, such as of distributions of magnetic properties, mixed magnetic phases\cite{Roberts2000}, clustering and long-range ferromagnetic state, magnetic characterisation  of geological mixtures and minerals and the differences in magnetization reversal mechanisms\cite{Pike2001,Muxworthy2005,Roberts2014,Gilbert2014}. The attractiveness of the FORC method is in its simplicity and its straightforward application to a wide range of systems displaying hysteresis. The accuracy of determining the SFD quantitatively in various classes of systems is presently under intensive critical discussion \cite{Dobrot??2013,Dobrot??2015}, and quantifying its range of validity, and understanding the microscopic reasons for its breakdown, is of broad interest with implications beyond the exemplar magnetic hysteresis considered here.

 In the simplest case of an assembly of bistable magnetic particles, the elementary hysteresis loop of a particle is rectangular and can be represented by a hysteron (Fig. \ref{fig1}a). In the absence of inter-particle interactions, when any magnetic correlations are irrelevant, the macroscopic hysteresis loop is simply a superposition of projections of magnetic moments of particles onto the field direction, ordered according to the switching events (hysteron thresholds) of individual particles. 
Then the SFD can be determined by de-constructing the hysteresis loop into distribution of hysterons uniquely linked with particle properties. In this case, FORC method is an inherently accurate technique for its identification. 
On the other hand, the presence of thermal relaxation and significant inter-particle interactions gives rise to more complex magnetisation reversal mechanisms (Fig. \ref{fig1}b). The emergent magnetic correlations fundamentally transform a macroscopic hysteresis loop and mask the direct information about the intrinsic switching fields of individual particles. The accuracy of the FORC method becomes parameter region-dependent and requires validation against the systematic inverse problem solving framework.

The purpose of the present article is to study the validity range of the FORC method against the large-scale computational data generated from a fully featured model of hard disk drive (HDD) media, which incorporates details of the statistical nature of inter-granular interactions, intrinsic properties of individual grains, and thermal activation.  Of particular importance is the role of magnetic correlations and consequent departures from simple hysteron-based model predictions. Gilbert et. al. \cite{Gilbert2014} have shown that the introduction of nearest-neighbour correlations strongly modifies the FORC diagram. Here we use a fully-featured kinetic Monte Carlo model with short- and long- ranged interactions to create a complete picture of interaction effects, most importantly the balance between exchange and magnetostatic interactions. We proceed with models of increasing complexity to demonstrate firstly the effects of thermal activation for a non-interacting system. We then proceed to the study of interaction effects using the kinetic Monte-Carlo (kMC) model and a simplified model of correlation effects. By evaluating the inter-granular magnetic correlation function, we demonstrate the direct relationship between the emergence of magnetic correlations and the failure of the FORC methodology to determine the SFD, and establish the criteria for the validity of the FORC method as a quantitative approach for accurate identification of the SFD in HDD magnetic media.

\section*{Results}


We apply the FORC method to large-scale computational data generated from a fully featured model of HDD media which incorporates details of the statistical nature of intergranular interactions, intrinsic properties of individual grains, and thermal activation (Methods \ref{methods_fullmodel}).
We use a kinetic Monte-Carlo (kMC) model (Methods \ref{methods_kmc}) to computationally reproduce the magnetisation behaviour of the HDD media, and a FORC method as the technique for identification of the underlying SFD.    
Here we consider realistic media with elongated grains with an aspect ratio ($h/d$) of 1.17, uniaxial anisotropy ($K$) with mean value of  $7 \cdot 10^6$ erg/cm$^3$ and 3 degree dispersion of the anisotropy easy axis around the perpendicular direction to the grain plane. The grain height ($h$) is 10 nm, the mean grain size ($d$) is 8.5 nm and the saturation magnetization ($M_s$) is $700$ emu/cm$^3$. The calculations assume an external field rate of $4 \cdot 10^4$ Oe/s at room temperature (300K). In all cases studied the intrinsic SFD can be easily calculated in the model by switching off all interactions (Methods \ref{methods_nonint}) and histogramming the switching fields of individual particles along a hysteresis loop. 

\subsection*{From reversal curve to FORC diagram and to SFD}

\begin{figure}[t!]
\begin{center}
\includegraphics[angle = 0,width = 11cm]{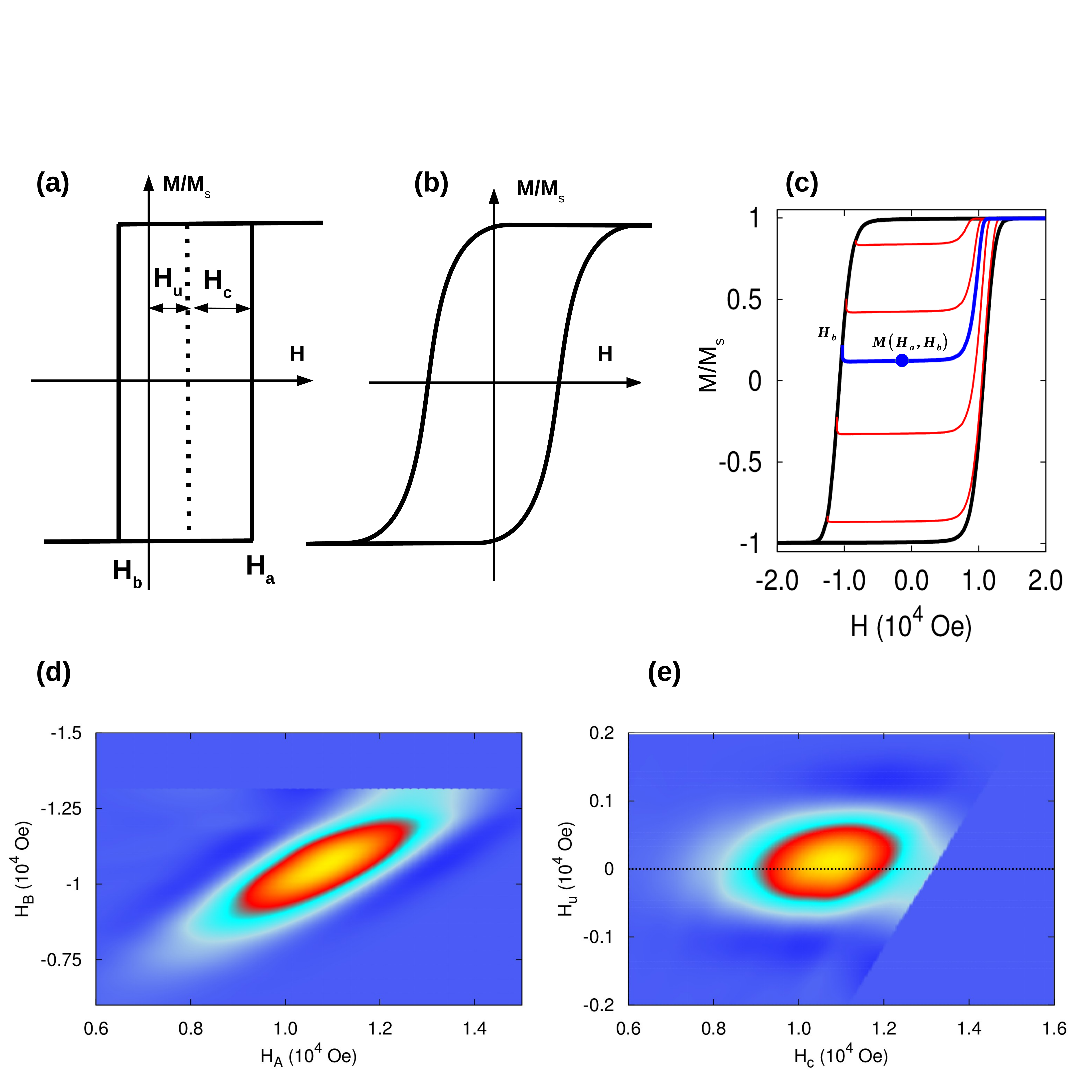}
\caption{ The ideal single particle hysteresis loop has a rectangular shape  corresponding to the hysteron model (a), where the only change in magnetisation is due to switching events.
The hysteresis loop will have a more complex shape (b) in the presence of thermal effects and reversal components as included in our model.
Example of hysteresis loops for a HDD media system (c) and the corresponding first order reversal curves  in $H_a H_b$ plane (d) or $H_c H_u$ plane (e) for a non-interacting system of 10000 elongated grains (1.17 aspect ratio and D=8.5nm) simulated at 300K and field rate of $4 \cdot 10^4$ Oe/s . The system parameters are: $M_s= 700$ $emu/cm^3$, K=$7 \cdot 10^6 erg/cm^3$ with 3 degree dispersion of the easy axis.  }
\label{fig1}
\end{center}
\end{figure}

The FORC method is used as a quantitative tool to investigate the SFD and interaction field distribution in granular materials. It is typically applied to the measurements of macroscopic hysteresis loops. The application of the method contains two main steps. The first step requires measurement of the first order reversal curves (FORC) and their transformation to the so-called FORC diagram (Fig. \ref{fig1} ). In the second step, the FORC diagram is processed such that the undesirable contribution of the inter-particle interaction is removed, which then allows accessing information about the intrinsic SFD.

\emph{FORC data, FORC diagram, and the SFD.} Fig. \ref{fig1}(c) illustrates the measurement protocol used to generate the FORC data. The starting point is the saturation of the sample by applying a large positive applied field. The field is then decreased towards the reversal field, $H_b$, when the field direction is reversed and increased from $H_b$ back to  positive saturation. This process generates a FORC attached to the major hysteresis loop at the reversal point $H_b$ (blue line in Fig. \ref{fig1}(c)). The magnetisation point at an applied field $H_a > H_b$ along this FORC, denoted as $M(H_a ,H_b)$, is internal to the major hysteresis loop. As illustrated in Fig. \ref{fig1}(c), at any value of $H_a$ in the hysteresis region, there is an entire family of such internal magnetisation points $M(H_a ,H_b)$ distinguished by the reversal field $H_b$ of their corresponding FORCs.
%
The FORC data are then analysed by computing the numerical second-order derivative of the functional dependence $M(H_a, H_b)$ with respect to the applied field $H_a$ and $H_b$:
\begin{align}
\rho_{ab}(H_a, H_b) = -\frac{1}{M_s}\frac{\partial^2 M(H_a, H_b)} {\partial H_a\partial H_b}
\label{forcab}
\end{align}
where $M_s$ is the saturation magnetisation of the material. It is next conventional to transform $\rho_{ab}$ by introducing new variables $H_c$ and $H_u$ such that $H_a(H_c, H_u) = (H_u + H_c)/2$ and $H_b(H_c, H_u) = (H_u - H_c)/2$, which leads to the FORC distribution represented as:
\begin{align}
\rho_{ab}(H_a, H_b) = \rho_{ab}(H_a(H_c, H_u), H_b(H_c, H_u)) \equiv \rho(H_c, H_u)
\label{forcuc}
\end{align}
from which the SFD can be obtained by a straightforward integration over the variable $H_u$:\cite{Zimanyi2006}
\begin{align}
\rho_{SFD}(H_c)=\int_{-\infty}^\infty \rho(H_c, H_u)\,dH_u
\label{sfd}
\end{align}
The interpretation of these equations is as follows. The distribution $\rho_{ab}$ in Eq. \eqref{forcab} is defined in terms of the differentiation of the magnetisation $M(H_a, H_b)$ attained through general applied fields $H_a$, $H_b$ along the hysteresis loop (Fig. \ref{fig1}(c)), and it is not immediately obvious how it relates to microscopic material properties such as the distribution of intrinsic switching field thresholds of magnetic grains $\rho_{SFD}$. The key to establishing this link is the notion of a magnetic particle having an elementary rectangular hysteresis loop (RHL) as shown in Fig. \ref{fig1}(a), with the up and down switching thresholds corresponding to the fields $H_a$ and $H_b$. Then, Eq. \eqref{forcab} can be interpreted as measuring the fraction of magnetic grains with the switching thresholds $H_a > H_b$ adding up to the cumulative magnetisation $M(H_a, H_b)$ at the field $H_a$ after the field excursion from $H_b$. The transformed variables $H_c = (H_a - H_b) / 2$ and $H_u = (H_a + H_b) / 2$ then represent the coercive and the bias fields of such RHLs (Fig. \ref{fig1}(a)), and the FORC distribution $\rho$ defined in Eq. \eqref{forcuc} is the joint probability distribution of $H_c$ and $H_u$. Consequently, the SFD defined by Eq. \eqref{sfd} is the distribution of the coercive fields of particles, i.e. their intrinsic switching thresholds.

In the ideal system of isolated magnetic particles represented by RHLs, such as the non-thermal system of non-interacting Stoner-Wohlfarth particles with the anisotropy axes aligned along the field direction, the RHLs have symmetric up and down switching thresholds $\pm H_c$ and due to the absence of interactions the $H_u=0$ for all particles. The macroscopic hysteresis loop is a superposition of magnetic states of all particles and, due to the rectangular shape of RHLs, any magnetisation change along the hysteresis loop can occur only at applied fields corresponding to the particle switching thresholds. The differentiation in Eq. \eqref{forcab} filters the contribution from the `flat' parts of RHLs and as a residual the distribution $\rho_{ab}$ carries an accurate representation of the switching thresholds of particles. In this case, the transformed FORC distribution in Eq. \eqref{forcuc} can be shown to be $\rho(H_c, H_u)=\rho^*(H_c)\delta(H_u)$, where $\delta(H_u)$ is the Dirac delta-function and $\rho^*(H_c)$ the statistical distribution of coercive fields of RHLs of particles, which according to Eq. \eqref{sfd} gives the SFD directly as $\rho_{SFD}=\rho^*(H_c)$.

Historically, an elementary RHL of a particle has been referred to as a hysteron in Preisach modelling,\cite{Mayergoyz1985,Dobrot??2015} which has served as a basis for developing the FORC method.\cite{Pike1999} The essence of Preisach models is to represent the macroscopic hysteresis loops of materials as a superposition of RHLs with the RHL threshold distribution, termed as a Preisach distribution, defined identically as the $\rho_{ab}$ in Eq. \eqref{forcab} (Methods \ref{methods_preisach}). The uniqueness of identification of the Preisach distribution has been shown to be guaranteed if the macroscopic magnetisation data satisfy the wiping-out and congruency properties \cite{Mayergoyz1985, Stancu2003}. Consequently, if the wiping-out and congruency properties are satisfied, the FORC distribution $\rho_{ab}$ is a valid and unique Preisach distribution. Unfortunately, the straightforward interpretation of Eqs. \eqref{forcab}-\eqref{sfd} as given above does not apply in realistic cases when the particles are represented by non-ideal RHLs, the inter-particle interactions are relevant, or in the presence of thermal fluctuations. Moreover, general systems with hysteresis do not always display the wiping-out and congruency properties, and the accuracy and uniqueness of the identification of SFD from the FORC distributions needs to be established with respect to the relevant physical picture and by independent measurement methodologies. Such cases are analysed in detail below.

\emph{Effects on imperfect RHLs on FORC diagram}.
To access the effects of deviations of elementary hysteresis loop of particles from the RHLs on the accuracy of determining the SFD, we applied the kMC model to study the hysteresis loop behaviour of a reduced system of isolated magnetic particles represented as Stoner-Wohlfarth particles (Methods \ref{methods_nonint} and \ref{methods_kmc}). The intrinsic magnetic properties of particles in the model were set to represent a typical magnetic recording medium (Methods \ref{methods_parameters}), including a 3$^\circ$ misalignment of the particle anisotropy easy axes around the applied field direction, and the driving field rate set to $10^4$ Oe/s of a typical experimental MOKE setup, which determined the extent of thermal activation. The inter-particle interactions were turned off.

Fig. \ref{fig1}(b) shows an example of the computed hysteresis loop of an ensemble of isolated particles, which clearly deviates from RHL behaviour (Fig. \ref{fig1}(a)). The rounding features are typical of a loop with strong component from thermal activation. The computed macroscopic hysteresis loop of a system of 10000 non-interacting particles with representative FORCs is shown in Fig. \ref{fig1}(c). The FORC diagrams $\rho_{ab}$ and $\rho$, obtained from this loop by applying Eqs. \eqref{forcab} and \eqref{forcuc}, are shown in Figs. \ref{fig1}(d) and (e). Note, that given the nature of their transformation, the FORC distributions $\rho_{ab}$ or $\rho$ are related by the 45$^\circ$ rotation of the $(H_a, H_b)$ coordinate plane. 
Fig. \ref{fig1}(e) shows that the FORC distribution $\rho$ is no-longer a straight line $\rho(H_c, H_u)=\rho^*(H_c)\delta(H_u)$ as in the case of a system of ideal non-interacting particles with RHLs discussed above, and instead has a significant $H_u$ component even if the inter-particle interactions are absent. This is due to the particle hysteresis loop rounding seen in Fig. \ref{fig1}(b), when the change of magnetisation along the macroscopic loop no longer occurs only at the switching thresholds of particles, as in the ideal RHL case, but in addition includes a smooth nonlinear component from the rounding effect. The magnetisation data transformation through Eq. \eqref{forcab} then convolutes the residual of the differentiation of this smooth component with the actual distribution of the switching thresholds of particles, which in the FORC diagram becomes manifested as a `fictitious' $H_u$ field distribution (Fig. \ref{fig1}(e)). This poses a difficulty in the interpretation of the FORC diagram, which appears to suggest the presence of interactions in the system of non-interacting particles.
Nevertheless, we find that evaluating the underlying SFD in Eq. \eqref{sfd} based on this FORC diagram actually yields the accurate SFD $-$ as a result of the reflection symmetry of the FORC diagram around the $H_c$ axis, when the $H_u$ component of the FORC distribution simply integrates to unity after factorisation of $\rho(h_c, h_u)$ in Eq. \eqref{sfd}. The slightly non-symmetric peak seen in Fig. \ref{fig1}(e), is often observed experimentally in systems with thermal activation.

\begin{figure}[t!]
\begin{center}
\includegraphics[angle = 0,width = 0.9\textwidth]{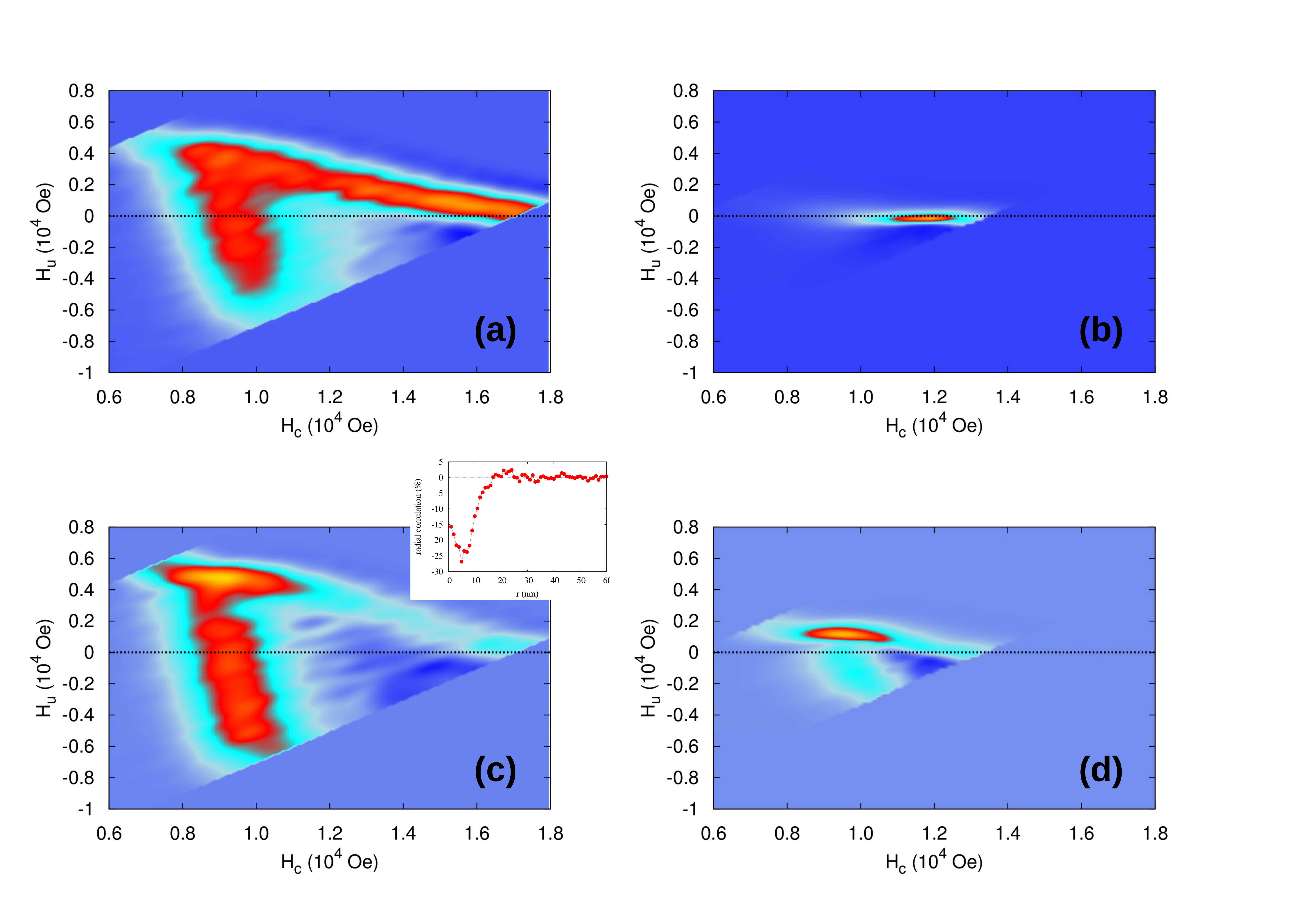}
\caption{ FORC diagram for interacting case with only magnetostatic interaction  calculated using mean-field approach (a) and grain-grain interaction  (c). The corresponding FORC diagram after mean interaction field is removed are given in (b) and (d) for the mean-field and grain-grain interaction models respectively. The radial magnetization correlation function for the grain-grain interaction model is given as the inset in (c).}
\label{fig2}
\end{center}
\end{figure}

\emph{Interactions: Mean-field correction of the FORC diagram.} The effects of inter-particle interactions on the FORC diagram are illustrated in Figs. \ref{fig2}, which shows that the FORC diagram becomes considerably modified by interactions with respect to the non-interacting case in Fig. \ref{fig1}(e). Fig. \ref{fig2}(a) shows the FORC diagram of the model based on the mean-field approximation of the full granular model (Methods \ref{methods_meanfield}). The mean-field interactions between grains have random strength following from a Gaussian distribution, obtained by histogramming the interaction fields of a full granular system in saturation used as a reference, weighted by the overall magnetisation $M$ of the granular system (Methods \ref{methods_meanfield}). The observed FORC diagram has the shape of a rotated `V' or `L' as expected \cite{Gilbert2014}. 

To apply the FORC method to identify the underlying SFD distribution, it is first necessary to extract the mean-field interaction and recover the non-interacting particle FORC diagram. This can be achieved by introducing a correction factor $\alpha$, variation of which allows to symmetrise the FORC diagram equivalently to subtracting the average interaction field acting on the system\cite{Papusoi2011}.
Specifically, varying the mean-field correction factor $\alpha$ transforms the field axes $H_a$ and $H_b$ in the raw FORC diagram (Fig. \ref{fig2}(a)) to new axes $H_a\rightarrow H_a-\alpha M(H_a)$ and $H_b\rightarrow H_b - \alpha M(H_b)$, until obtaining the optimal value of $\alpha\equiv\alpha_o$ when the new FORC diagram $\rho$ becomes symmetric around the $H_u$ axis and any possible negative regions of $\rho < 0$ that may result from an over- or under-estimated mean-field correction become eliminated. This procedure is equivalent to the hysteresis loop `de-shearing' procedure typically applied to extract the effects of demagnetising fields from experimental hysteresis loops. In the ideal case, the optimal value of the correction factor, $\alpha_o$, corresponds to the mean-field interaction strength $\langle H_{inter}\rangle$ of the mean field granular model (Methods \ref{methods_meanfield}). After applying the mean-field correction, the resulting FORC diagram shown in Fig. \ref{fig2}(b) resembles that of the non-interacting case  Fig. \ref{fig1}(e), which allows to calculate the SFD  by applying Eq. \eqref{sfd}. The values found are consistent with the non-interacting cases within the statistical error corresponding to uncertainty of 5\%.

\emph{Interactions: magnetic clusters.} The mean-field interaction is expected to be an oversimplification as it does not  account for the inter-granular magnetic correlations typically present in real systems. The presence of such correlations leads to the emergence of magnetic clusters which influences the accuracy of the FORC method. To begin investigating the magnetic clustering effect, we first consider the mean-field model discussed above reduced to an ensemble of disconnected regions of $N_g$ grains.  In this `toy model' the regions act as non-interacting clusters of $N_g$ grains interacting via equivalent mean-field-like interactions dependent on the average magnetisation within each cluster (Methods \ref{methods_clusters}). In this model, a switching grain affects only the magnetisation of its own cluster, while the magnetisation of all other clusters in the ensemble remains unaffected, and the hysteresis loop is a superposition of magnetisation jumps from individual clusters. Thus, when the cluster size $N_g$ is large, approaching the system size, the behaviour recovers that of a full mean-field system discussed above. On the other hand, as $N_g$ decreases the behaviour moves away from being mean-field-like and the macroscopic loop results from a combined contribution of an increasing number of elementary hysteresis loops of individual clusters available in the system. These elementary loops of individual clusters have shape deviating from the RHLs, which is expected to reduce the accuracy of the FORC method. 

\begin{figure}[h!]
\begin{center}
\includegraphics[angle = 0,width = 0.7\textwidth,scale=0.4]{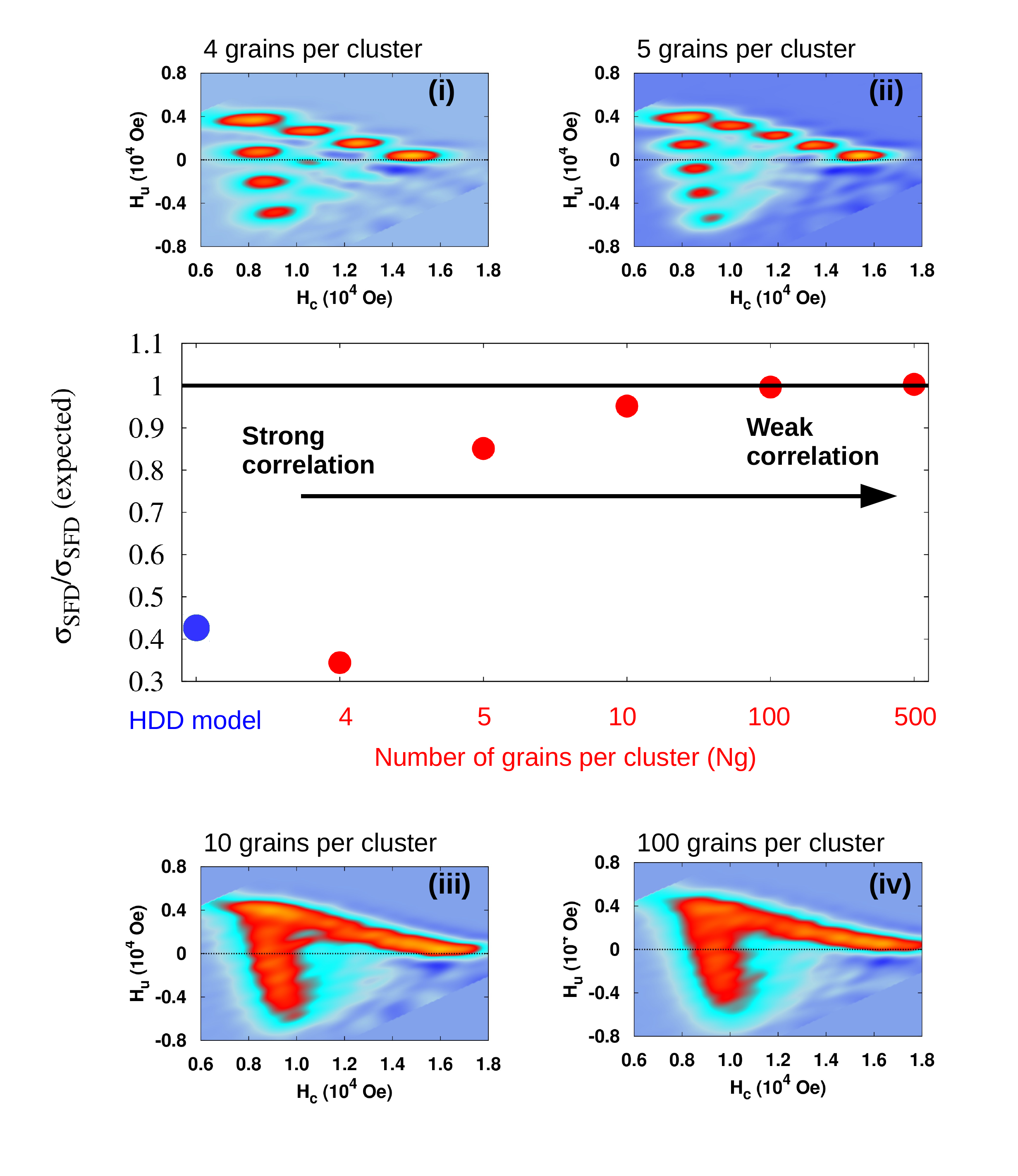}
\caption{ The width of the SFD ($\sigma_{SFD}$) as function of cluster size for the toy model (red). The cluster size in inverse proportional with the degree of correlation in the system. The larger the cluster size the closer is the model to a mean-field-like model, which is completely uncorrelated system and FORC method can be applied successfully. By increasing the correlation, the FORC method is underestimating the $\sigma_{SFD}$.  The result from the HDD model is also included (blue). Example of FORC diagram for different cluster size are illustrated in the insets i-iv. }
\label{fig22}
\end{center}
\end{figure}

Fig. \ref{fig22} shows analysis of five ensembles of uniform clusters of variable $N_g=4$, 5, 10, 100, 500. Applying the cluster model (Methods \ref{methods_clusters}) combined with the kinetic Monte-Carlo solver (Methods \ref{methods_kmc}) we first computed the macroscopic hysteresis loops with FORCs for every ensemble. Then we transformed the FORC data to the underlying FORC diagram by applying Eqs. \eqref{forcab} and \eqref{forcuc}, applied the mean-field correction $\alpha_0$ to remove the interactions as discussed above - which is a standard procedure used in the practical FORC method, and computed the SFD from the corrected FORC diagram using Eq. \eqref{sfd}. As expected, the results of extracting the SFD are accurate when $N_g$ approaches the full system size, while the accuracy of the FORC method reduces with the decreasing cluster size $N_g$. When $N_g$ is small, the clusters contain only small numbers of grains relative to the full system size, and there are many clusters contributing to the overall macroscopic hysteresis loop. 
There are two main sources of error expected to contribute to the loss of accuracy of the FORC method: 1) the mean-field correction is no-longer accurate as in the mean-field model of a full granular system, and 2) distorted RHL shape of elementary hysteresis loops of individual grains, due to correlated behaviour inside each cluster.
Examples of the obtained raw FORC diagrams prior applying the mean-field correction are shown in the insets i-iv in Fig. \ref{fig22}. The mean-field like nature of the interaction in the cluster model (Methods \ref{methods_clusters}) results in an equivalent effective shift of the switching thresholds of the grains in each cluster, which results in the observed segmentation of the `V'-like shape FORC diagram into distinct regions along each branch. The number of these regions per branch corresponds to the number of grains per cluster, the `V' shape of the arrangement of the segments reflects the interaction induced symmetry breaking of the up and down intrinsic switching thresholds of grains where the separation between the segments corresponds roughly to the magnitude of the mean interaction field $\langle H_{inter}\rangle$ in the model. The interpretation of this FORC diagram is consistent with the recent work, where analogous segmentation effects have been studied in terms of a different model with the nearest neighbour grain interactions \cite{Gilbert2014}. 
Increasing $N_g$ in clusters results in the increased density of segments in the FORC diagram until gradually reproducing the FORC diagram of the mean field model in Fig. \ref{fig2}(a).

\emph{Applicability of the FORC method to a full granular model of recording media.}
To investigate the accuracy of the FORC method in determining the SFD in realistic magnetic recording materials we use the full granular kinetic Monte-Carlo model with exchange and magnetostatic interactions (Methods \ref{methods_fullmodel}) to simulate the underlying hysteresis loops and FORCs. Such general interactions introduce magnetic correlations between grains, which lead to correlated behaviour when magnetic grains begin to switch in unison in clusters of size equal to the characteristic correlation length. This leads to magnetisation jumps (Barkhausen noise) along the hysteresis loop, analogous to the case of the cluster model discussed above.
Fig. \ref{fig2}(c) shows the corresponding FORC diagram. The inset in the figure shows the radial correlation function, suggesting the presence of significant short range grain-grain correlations in a typical recording medium which are absent in the non-interacting and mean-field granular systems. To remove the contribution from interactions, we first subtract the mean-field correction after finding the optimal $\alpha_o$ as discussed above, as is typically done in practical applications of the FORC method. The corrected FORC diagram shown in Fig. \ref{fig2}(d) deviates from the non-interacting case shown in Fig. \ref{fig1}(e), which is due to the fact that the mean-field interaction mis-represents the full exchange and magnetostatic interactions. Consequently, applying Eq. \eqref{sfd} we find that the FORC method underestimates the $\sigma_{SFD}$ by as much as 60\%. Thus the presence of significant magnetic correlations results in the loss of accuracy of the FORC method. The question of main interest is to understand the relationship between the extent of correlations and the accuracy of the SFD determined by the FORC method. 

\begin{figure}[t!]
\begin{center}
\includegraphics[angle = -0,width = 0.7\textwidth]{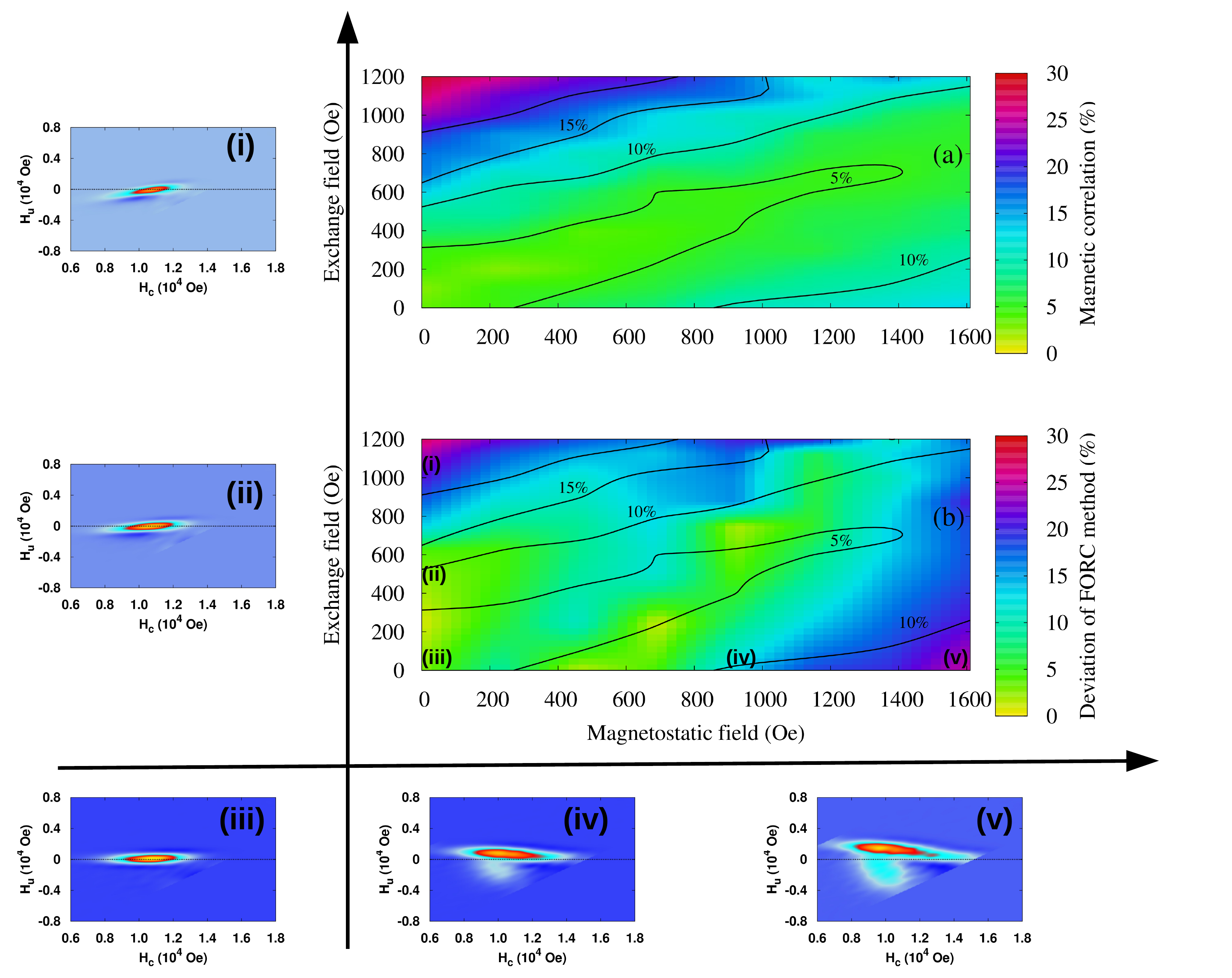}
\caption{ Plots of correlation and error from the FORC calculations;
(a) Correlation diagram: The magnetization correlation is calculated at coercivity and the maximum correlation is extracted. As exchange and magnetostatic interactions increase, the coupling between grains also increases leading to large correlation values. The values on the diagonal are minimum because positive and negative contributions from the  exchange and magnetostatic interaction, compensate overall.
 (b) Validity diagram: Diagram showing the deviation of $\sigma_{SFD}$ from the FORC method in comparison with the expected value. The contour lines for different correlation in (a) are used in (b) to guide the eye.
 (inset i-v) Example of FORC diagram for the system having just exchange interaction (ii: 500 Oe and i: 1125 Oe), having just magnetostatic interaction (iv: 920 Oe, V: 1600 Oe). The non interacting FORC diagram is illustrated for comparison in inset iii.
}
\label{fig_Diagram}
\end{center}
\end{figure}

To study this issue in simulations, we systematically varied the strength of exchange and magnetostatic interactions, in each case computing the underlying radial pair correlation function between the grains (Methods \ref{methods_correlations}), and evaluated the reference SFD directly by histogramming the intrinsic field thresholds of grains during switching for comparison with the SFD obtained by the FORC method through Eqs. \eqref{forcab}-\eqref{sfd}.
Fig. \ref{fig_Diagram}(a) shows the dependence of the maximum value of the correlation function on the strength of exchange and magnetostatic fields. Representative FORC diagrams after applying the mean-field correction are shown in the insets (i)-(v). Magnetic correlations increase with the strength of one of the interaction types increasing relative to the other, while they remain negligible in the weakly interacting case or in the interaction compensating region corresponding to the region with similar total magnitudes of exchange and magnetostatic interactions. The contour lines quantify the correlation strength. Fig. \ref{fig_Diagram}(b) shows the corresponding relative accuracy of the SFD determined by the FORC method, measured relative to the SFD determined directly from the kMC model.
The comparison of Figs. \ref{fig_Diagram}(a) and (b) reveals close agreement between the correlation strength and the accuracy of the FORC method for determining the SFD. Errors can also be attributed to the fact that the model is thermal and RHL are not perfect, or the effects of the slight misalignment of anisotropy axes of grains combined with the interactions. However, as shown in Fig. \ref{fig1}(e) for the thermal effects, these factors are relatively small and the largest discrepancy is caused by the interactions and specifically by the interaction-induced spatial correlations. The accuracy of the FORC method is the highest in the weakly correlated interaction regions.  
Nevertheless, depending on the required accuracy of determination of the SFD, Fig. \ref{fig_Diagram}(b) indicates the range of parameter space in which this can be achieved. The FORC method is limited to very small field (up to 1200Oe) considering a deviation of 10\% from the expected value of $\sigma_{SFD}$. Finally we map the deviation of the SFD from FORC and combine the results with magnetization correlation data to draw a validity diagram for using FORC as a quantitative tool.  

\section*{Discussion}

Our study presents the analysis of the accuracy and the different modes of failure of the FORC method by using as a benchmark a succession of computational models of increased complexity starting from the system of non-interacting particles towards the realistic full model of magnetic granular media in magnetic recording, which includes exchange and magnetostatic interactions and various relevant sources of the material disorder.
In terms of the analysis, similar to Pike et.al. \cite{Pike1999} we make the distinction between the raw FORC data, the FORC diagram and the usual interpretation of the FORC data based on the Preisach model. Using model calculations we show that, while the FORC diagram in principle contains information about the SFD and the interactions, application of the RHL interpretation does not reliably deconvolve the SFD and interaction effects. This is attributed to the spatial magnetization correlations which are an important feature of many materials, including magnetic recording media, and which are not included in the RHL approach. 

We reveal that the applicability of the FORC method for the quantitative analysis of the SFD is limited to the parameter range where inter-particle spatial correlations are insignificant, i.e. when exchange and magnetostatic interactions are weak or when they compensate. 
The accuracy of the FORC method decreases in the presence of significant correlations resulting from the correlated switching with multiple grains reversing their magnetic state in unison at a given field threshold. These correlated grains behave as a single entity thus hiding any information about the intrinsic switching thresholds of individual particles into the correlated reversal. The information cannot be recovered by the differentiation of the first order magnetisation reversal curves in the way of the FORC method simply because Eq. \eqref{forcab} no longer provides access to the switching thresholds of individual RHLs of particles but instead provides access to the fields corresponding to the magnetisation jumps along first order reversal curves, which are equivalent to the switching thresholds of the correlated particle clusters as a single entity and not as individual particles in the cluster. 

Moreover, general hysteresis loops do not satisfy the wiping out and congruency properties. Then the FORC diagrams cannot be interpreted as Preisach distributions and no longer guaranteeing unique SFD \cite{Dobrot??2015, Mayergoyz1985, Pike1999,Dobrot??2013}, which necessitate careful analysis in establishing its physical relevance.  
Recovering the intrinsic switching thresholds of individual particles from the first order reversal curve data then requires further deconvolution based on the refined models capable of accounting for the detailed structure of inter-particle interactions. Our work applies such fine-scale models and based on the evaluation of microscopic correlation functions establishes quantitatively the range of validity of the FORC method for determining the SFD with relevance to magnetic recording media (Fig. \ref{fig_Diagram}).

Generally speaking, to identify the SFD in the material parameter range beyond the applicability of the FORC method requires inverse problem solving techniques based on the physically realistic models, which allow reproducing the relevant correlated switching of particles. Moreover, besides identifying accurate models suitable for interpreting the experimental data, such methods also require establishing uniqueness properties of the identified solutions. We have implemented a direct approach employing optimisation techniques based on the grid-search method \cite{Hovorka2009} to fit the full recoding model (Methods \ref{methods_fullmodel}) to the computed hysteresis loop data, and uniquely recovered the expected SFD in the entire parameter range.
Thus, the most reliable, albeit computationally expensive approach, seems to be to essentially carry out by a direct fit to the experimental FORC data using a microscopic approach, including the detailed calculation of the interactions such as presented here for the specific example of perpendicular recording media.
%
 %
 
\section*{Methods}

\subsection{Full interacting model recording media}\label{methods_fullmodel}
The system consists of $N$ Stoner-Wohlfarth  grains, where the volume ($V$) and geometry of the grains is generated by using a Voronoi construction. The energy of a system of $N$ grains is:
\begin{equation}\label{energy}
E = \sum_iK_i V_i(\hat k_i\times\hat m_i)^2
- \sum_{i} M_sV_i\hat m_i\cdot\vec {H}_{ap}
- \frac{1}{2}\sum_{nn\,\,ij}E^{ij}_{exch}
- \frac{1}{2}\sum_{i\ne j} E^{ij}_{mag}
\end{equation}
where the first term is the uniaxial anisotropy terms with $\vec K_i = K_i\hat k_i$ being the uniaxial anisotropy vector and $V_i$ the volume of a particle $i$, $M_s$ the saturation magnetisation, and  $\hat m_i = \vec m_i/M_s$ the particle moment normalised to unity. The values of $K_i$, $\hat k_i$, and $V_i$ are drawn from random distributions relevant to modern granular magnetic recording materials, as described below. The second term represents the Zeeman term describing the interaction of grains with the applied field $\vec H_{ap}$.

The third term in Eq. \eqref{energy} describes the exchange interaction between the nearest neighbour grains. The exchange interaction in granular materials for magnetic recording is dependent on the extent of the grain boundary and is of randomised character, which can be expressed as $E^{ij}_{exch} = M_sV_i\hat m_i\cdot H^{ij}_{exch}$ with the locally varying exchange field $H^{ij}_{exch}$:\cite{Peng2011}
\begin{equation}
H^{i,j}_{exch}=H_{exch} \left( \frac{J_{ij}}{\langle J_{ij}\rangle} \right) \left( \frac{L_{ij}}{\langle L_{ij}\rangle} \right) \left( \frac{\langle A_{i}\rangle}{A_{i}} \right)
\label{exch}
\end{equation} 
where $H_{exch}$ is the mean strength of the exchange interaction field, $J_{ij}$ is the fractional exchange constant between the adjacent grains $i$ and $j$ with $L_{ij}$ being the length of the connecting boundary, $A_i$ is the area of the grain $i$, and $\langle \cdot\rangle$ represent averages over all pairs of grains. 

The last term in Eq. \eqref{energy} represents the magneto-static interaction between the grains and is represented as $E_{mag}^{ij} = M_sV_i\hat m_i\cdot H_{mag}^{ij}$.
The contribution to the magneto-static interaction field $H_{mag}^{ij}$  is performed by a direct integration of the magneto-static surface charge \cite{Newell1993}. The evaluation of $H_{mag}^{ij}$ by full integration over the surface charge accounts for the correction resulting from the dipolar interaction over-estimating the magneto-static interaction in the proximity of a grain.\cite{Liu2009a} Both exchange and magnetostatic interactions in Eqs. \eqref{energy} are dependent on the size and shape of grains, and on the inter-granular distance.

\subsection{Approximations of the full model}
Various levels of reduction of the full model can be introduced as follows. 

\subsubsection{Non-interacting model approximation of recoding media}\label{methods_nonint}

In the non-interacting model, the definition of the system energy reduces from Eq. \eqref{energy} to:
\begin{equation}\label{energy0int}
E = \sum_iK_i V_i(\hat k_i\times\hat m_i)^2
- \sum_{i} M_sV_i\hat m_i\cdot\vec {H}_{ap}
\end{equation}
The kinetic Monte-Carlo modelling of this system allows to study thermal relaxation aspects in ensemble of non-interacting Stoner-Wohlfarth particles, and may serve as a reference for gauging the effects of interactions in the full interacting model.

\subsubsection{Mean-field model of recoding media}\label{methods_meanfield}
In the mean-field model the interactions between magnetic grains are introduced in a uniform ways. The energy expression given in Eq. \eqref{energy} can be reduced:
\begin{equation}\label{energymf}
E = \sum_iK_i V_i(\hat k_i\times\hat m_i)^2
- \sum_{i} M_sV_i\hat m_i\cdot\vec {H}_{ap}
- \sum_i M_sV_iH^i_{inter} \langle\hat m_k\rangle\cdot\hat m_i
\end{equation}
where the symbol $\langle\hat m_k\rangle$ implies averaging over all grains in the system, i.e. average magnetisation at a given field $\vec {H}_{ap}$, $H_{inter}^i$ is the random interaction field given by Gaussian distribution with mean $\langle H_{inter}\rangle$ and standard deviation $\sigma_{inter}$. We found that Gaussian distribution represents well the distribution of interaction fields in the full HDD model at saturating fields. This allows us to calibrate the mean-field interaction strength to be consistent with the full model at saturating fields, which is a point used as a reference.

\subsubsection{Cluster ensemble model of recording media}\label{methods_clusters}
In the cluster model, the full model is divided into clusters of $N_g$ grains, with grains inside a $j$-th cluster interacting via a mean-field like interaction, while the clusters being non-interacting. The full energy expression given in Eq. \eqref{energy} reduces to a sum through individual clusters $j$ as $E = \sum_{j} E_j$ with:
\begin{equation}\label{energycluster}
E_j = \sum_{i\in j}^{N_g}K_i V_i(\hat k_i\times\hat m_i)^2
- \sum_{i\in j} M_sV_i\hat m_i\cdot\vec {H}_{ap}
- \sum_{i\in j} M_sV_iH^i_{inter} \langle\hat m\rangle_j\cdot\hat m_i
\end{equation}
where the symbol $\sum_{i\in j}$ implies that the summations occur through the magnetic grains with the cluster $j$, and $\langle\hat m_k\rangle_j$ is the average magnetisation of the cluster $j$. The interaction field $H_{inter}^i$ is defined identically as in the mean-field case in Section \ref{methods_meanfield}.

\subsection{Modelling hysteresis with thermal activation: Kinetic Monte-Carlo approach}\label{methods_kmc}
The thermal fluctuation and external field driven magnetisation behaviour of interacting magnetic particles as described in the model in the Methods Section 0.1 is modelled by using kinetic Monte-Carlo approach\cite{Chantrell2000,Ruta2015}. The effective local fields of particles are given by Eqs. \eqref{energy} as $\vec {H}_{ap}+\vec{H}^{ij}_{mag} +\vec{H}^{ij}_{exch}$. The time dependent transition for a particle moment $\hat m_i$ to switch between the up (`1') and down (`2') states is $P_i = 1-\exp(-t/\tau_i)$, where the relaxation time constant $\tau_i$ is a reciprocal sum of the transition rates $\tau^+_i$ and $\tau^-_i$ dependent on the energy barriers $\Delta E_i^{1,2}$ seen from the `1' and `2' states via the standard N\'eel-Arrhenius law\cite{Neel1949}: $\tau_i^{1,2} = \tau_0\exp(\Delta E_i^{1,2}/k_BT)$. The $k_B$ is the Boltzmann constant and $T$ the temperature. According Eq. \eqref{energy}, the $\Delta E_i^{1,2}$ depend on the intrinsic particle properties, such as $V_i$ and $\vec K_i$. 

\subsection{Simulation parameters of realistic recording media}\label{methods_parameters}

Throughout this study we consider a thin film system, with elongated grains (1.17 aspect ratio), log-normal volume distribution (33\%) and log-normal anisotropy distribution (5\%). The uniaxial anisotropy has a 3$^\circ$ dispersion of easy axis around the axis perpendicular to the film. 
The system properties are: mean anisotropy $\langle K_i\rangle=7 \cdot 10^6$ erg/cm$^3$, saturation magnetisation $M_s = 700$ emu/cm$^3$, grain height $h = 10$ nm and the mean grain size $d = 8.5$ nm. The calculations are done for an external field rate of $4 \cdot 10^4$ Oe/s at room temperature 300K.

\subsection{Rectangular hysteresis loop (RHL) model: Preisach modelling}\label{methods_preisach}
If a granular system can be viewed as a collection of grains having rectangular hysteresis loops (RHL), with coercive field $H_c$ and bias field $H_u$ given by probability distribution, then the macroscopic hysteresis loop of the system can be obtained as a superposition of the RHLs and magnetisation $M(H_a, H_b)$ represented as:
\begin{equation}\label{preisach}
M(H_a, H_b) / M_s = 
\int_0^{AB}dH_c \int_{-A}^{A}\rho(H_c, H_u)dH_u +
\int_{AB}^\infty dH_c \int_{-B}^{B}\rho(H_c, H_u)dH_u
\end{equation}
where $M_s$ is the saturation magnetisation, $AB = (H_a - H_b)/2$, $A = H_a - H_c$, and $B = H_b + H_c$ are the integration limits dependent on the applied field $H_a$ along the FORC attached do decreasing major hysteresis loop at the reversal field $H_b$, i.e. $H_a > H_b$.
Applying the Leibniz integral rule to differentiate the integral we obtain:
\begin{equation}\label{forc1}
\frac{1}{M_s}\frac{\partial^2 M(H_a, H_b)}{\partial H_a\partial H_b} = -\rho\left(\frac{H_a-H_b}{2}, \frac{H_a+H_b}{2}\right)
\end{equation}
Given that Eq. \eqref{preisach} is inherently a superposition from switching events of individual grains, Eq. \eqref{forc1} establishes the relation between the applied fields $H_a$ and $H_b$, and the intrinsic switching thresholds of particles, which can be labeled equivalently as $H_a$ (threshold of a grain flipping up along the FORC at the field $H_a$) and $H_b$ (threshold for a grain flipping down before generating FORC at the field $H_b$). Given that $H_c = (H_a - H_b)/2$ and $H_u = (H_a + H_b)/2$ (Fig. \ref{fig1}(a)), the above equation can be rewritten as:
\begin{equation}
\rho(H_c(H_a, H_b), H_u(H_a, H_b))\equiv\rho_{ab}(H_a, H_b) = 
-\frac{1}{M_s}\frac{\partial^2 M(H_a, H_b)}{\partial H_a\partial H_b}
\end{equation}
which agrees with the definition of the FORC distribution given in Eqs. \eqref{forcab} and \eqref{forcuc}. If the system displays the wiping-out and congruency properties, Eq. \eqref{preisach} can be shown to be a unique Preisach distribution associated with the granular system represented by magnetisation $M(H_a, H_b)$.

\subsection{Magnetic correlation}\label{methods_correlations}
To investigate the coupling between grains due to correlated behaviour, we computed the  radial correlation function as following:
\begin{align}
C_j(r)=\frac{\langle m_j(R) m_j(R + r)\rangle - \langle m_j(R)\rangle\langle m_j(R + r)\rangle}{\sqrt{\langle m_j^2(R)\rangle - \langle m_j(R)\rangle^2 }\sqrt{\langle m_j^2(R+r)\rangle - \langle m_j(R+r)\rangle^2 } } ,
\end{align}
where $j=x,y,z$ and $m_j(R)$, $m_j(R+r)$ are pairs of of grains separated by a distance $r$. The correlation data plot in Fig. \ref{fig_Diagram}(b) shows the correlation function $C_z(r)$.

\subsection{FORC method}
The measurement protocol to produce a first order reversal curve (FORC) begins by first applying a large field to saturate the sample, then decreasing the field to a certain value $H_b$. From this point, the FORC is obtained by increasing the field back to saturation. The magnetisation is recoded at fields $H_a$ along the FORC at the reversal field $H_b$, $H_a>H_b$. The FORC diagram is then evaluated using Eqs. \eqref{forcab} and \eqref{forcuc}, from which the SFD can be calculated using Eq. \eqref{sfd}.

\section*{Acknowledgements }

We are grateful to Prof. A. Berger,  Prof. A. Stancu,  Prof. G.T. Zimanyi, M. Strungaru and A. Meo  for helpful discussions and comments. This work made use of the facilities of N8 HPC provided and funded by the N8 consortium and EPSRC (Grant
No. EP/K000225/1) co-ordinated by the Universities of Leeds and Manchester and the EPSRC Small items of research equipment at the University of York ENERGY (Grant No. EP/K031589/1).

\end{document}